\documentclass[12pt,english]{article}
\usepackage[T1]{fontenc}
\usepackage[latin9]{inputenc}
\usepackage{geometry}
\usepackage{graphicx}
\usepackage{amssymb}

\usepackage{babel}

\begin{document}

\title{Conversion of graded to binary response in an activator-repressor
system}

\author{Rajesh Karmakar%
\thanks{Electronic address: rkarmakar2001@yahoo.com%
}}

\maketitle
\begin{center}
Department of Physics 
\par\end{center}

\begin{center}
A. K. P. C. Mahavidyalaya 
\par\end{center}

\begin{center}
Subhasnagar, Bengai, Hooghly-712 611, India. 
\par\end{center}
\begin{abstract}
\noindent Appropriate regulation of gene expression is essential to
ensure that protein synthesis occurs in a selective manner. The control
of transcription is the most dominant type of regulation mediated
by a complex of molecules such as transcription factors. In general,
regulatory molecules are of two types: activator and repressor. Activators
promote the initiation of transcription whereas repressors inhibit
transcription. In many cases, they regulate the gene transcription
on binding the promoter mutually exclusively and the observed gene
expression response is either graded or binary. In experiments, the
gene expression response is quantified by the amount of proteins produced
on varying the concentration of an external inducer molecules in the
cell. In this paper, we study a gene regulatory network where activators
and repressors both bind the same promoter mutually exclusively. The
network is modeled by assuming that the gene can be in three possible
states: repressed, unregulated and active. An exact analytical expression
for the steady-state probability distribution of protein levels is
then derived. The exact result helps to explain the experimental observations
that in the presence of activator molecules the response is graded
at all inducer levels whereas in the presence of both activator and
repressor molecules, the response is graded at low and high inducer
levels and binary at an intermediate inducer level. 
\end{abstract}
\noindent \begin{flushright}
PACS number(s): 87.10.Mn 
\par\end{flushright}

\textbf{\large I. INTRODUCTION}{\large \par}

\vspace{0.5cm}

Gene expression, a fundamental cellular process whereby mRNAs and
proteins are synthesized, is inherently stochastic in nature. There
is a large number of theoretical and experimental studies which confirm
the stochastic nature of gene expression \cite{Raj Review}. The stochasticity
or noise in gene expression is due to the small number of molecules
involved in the associated cellular processes. For example, the DNA
molecule which gives an organism its unique genetic identity is present
in one or two copies per cell. The small number of molecules taking
part in the biochemical events of gene expression is responsible for
the probabilistic occurrence of the events. The stochastic nature
of the biochemical events introduces fluctuations around the mean
mRNA and protein levels. The fluctuations constitute noise and cause
identical copies of a gene to express at different levels. The total
noise in the gene expression level has two components: intrinsic and
extrinsic. The origin of intrinsic noise lies in the probabilistic
nature of the biochemical events of gene expression. The sources of
extrinsic noise is in the fluctuations in cellular components such
as RNAPs, ribosomes and regulatory molecules. The noise in gene expression
may give rise to heterogeneity in a cell population. Cell-to-cell
variability is generally attributed to genetic differences though
the environment and history are also contributing factors. Recent
experiments \cite{Raser_Shea,Blake 2003} provide evidence that stochasticity
in gene expression can contribute substantially to population heterogeneity
and consequent variability in the cellular phenotype. A population
of cells with identical genetic sequences as well as history and subjected
to the same constant environment can develop heterogeneities due to
the random nature of gene expression. Cellular heterogeneity has been
observed in a variety of cell types ranging from bacteria \cite{Ozbudak }
to complex mammalian cells \cite{Ramsey}. Several experiments combined
with theoretical studies provide important new insight on the stochastic
aspects of gene expression \cite{Raser_Shea,Blake 2003,Ozbudak ,Thattai,Swain PS,Kaern Review}.

Gene expression and its regulation are of fundamental importance in
living organisms. There are many steps in gene expression pathway
from DNA to proteins and different types of regulatory molecules are
involved in different steps. In general, transcriptional regulation
is one of the most dominant types of regulation. The activator and
repressor molecules are actively involved in the regulation of gene
transcription both in prokaryotes and eukaryotes. Transcriptional
repressors such as lac and tryptophan repressors are well known for
prokaryotic systems. Repressor molecules inhibit the gene transcription
by binding to the appropriate region of the DNA. Eukaryotic systems
are much more complex and have compact chromatin structures. For the
initiation of transcription, remodelling of the chromatin structure
is essential so that the transcription factors and the RNA polymerase
have access to the appropriate binding regions. Thus, gene activation
in eukaryotic system means the relief of repression by the nucleosomal
structure of the chromatin. After remodelling of chromatin structure,
activator protein binds the DNA and activate gene expression. Activator
protein concentrations can be varied by varying the inducer molecules
such as galactose \cite{Blake 2003}.

Experiments reveal that in an individual cell the gene expression
response, the amount of proteins synthesized, can be of two types:
graded and binary. In graded response protein level varies continuously
with varying concentration of external inducer molecules. In binary
response, protein levels can have two possible values: low or high.
This is also known as the all-or-none phenomenon in gene expression.
The binary response at the single cell level gives rise to a bimodal
distribution in protein levels at the population level. There are
experimental evidences of binary responses in gene expression with
different possible origins \cite{Blake 2003,Becskei,Tan,Rossi,Bigger,Zlokarnik}.
Becskei et al. \cite{Becskei} have demonstrated that positive feedback
with cooperativity can generate binary response in a synthetic eukaryotic
gene circuit but without the positive feedback the response is graded.
The presence of positive feedback loop with cooperativity gives rise
to bistability and the bistability along with stochasticity produces
binary response in protein levels. Recently, Tan et al. \cite{Tan}
have established that bistability may also arises from the interplay
between a non-cooperative positive feedback loop and circuit-induced
growth retardation. Blake et al. \cite{Blake 2003} and Karmakar and
Bose \cite{Karmakar TF} have shown that fluctuations in the levels
of transcription factor can give rise to binary responses in the target
gene expression in an eukaryotic system. Rossi et al. \cite{Rossi}
and Biggar and Carbtree \cite{Bigger} have further shown that, in
certain instances, competition between activator and repressor molecules
to occupy the promoter region can generate a binary response in gene
expression. If the activator or repressor molecules act independently,
a graded response is obtained. The difference in the cellular fates
in binary response may be ascribed to heterogeneity in the distribution
of the stimulus/inducer molecules in the cell population, different
histories, i.e., initial states in the case of bistability, intracellular
noise giving rise to fluctuations in key parameter values etc.

To explain the experimentally observed binary responses in gene expression,
different modeling approaches have already been proposed and analyzed
using simulation and analytical techniques. Binary response in an
autocatalytic induction circuit is very common and easily understood
from different theoretical studies \cite{Carrier,Rajesh Comp}. Noise
can have important role in the generation of binary responses in gene
expression. It may be of purely stochastic origin \cite{Kepler,Piron,Karmakar}.
Kepler and Elston \cite{Kepler} have demonstrated through specific
examples that only stochasticity in gene expression can give rise
to binary response, i.e., a bimodal distribution in the protein levels.
Pirone and Elston \cite{Piron} show that the slow promoter transition
in gene states is responsible for binary responses whereas fast transitions
produce graded responses. Karmakar and Bose \cite{Karmakar} defined
the slow and fast transitions between the active and inactive states
of the gene more precisely and established the conditions of origin
of graded and binary responses in gene expression. They derived the
distribution of protein levels assuming the random transitions between
the gene states with protein synthesis and degradation occurring deterministically.
Later, exact analytical distributions for mRNAs and proteins have
been derived considering all the major steps of gene expression i.e.,
transcription, translation and degradation, to be stochastic \cite{Raj mRNA,Swain exact}.
In this paper, we propose a simple model of stochastic gene transcription
regulated by activators and repressors and show using exact analytical
calculations that bimodal distribution in protein levels appears naturally
when activators and repressors compete for the binding site mutually
exclusively to regulate the gene transcription. On the other hand,
a graded response is observed when only activator molecules regulate
the gene transcription.

\vspace{0.5cm}

\textbf{\large II. STOCHASTIC MODEL AND EXACT SOLUTION}{\large \par}

\vspace{0.5cm}

Transcriptional regulation by activator and repressor molecules on
binding the same promoter is an important regulatory mechanism of
gene expression in living organisms. The activator (repressor) molecules
activate (inhibit) the transcription by binding the appropriate site
on the promoter. Here we consider a gene regulatory network where
activators and repressors both regulate the gene transcription mutually
exclusively \cite{Rossi}. This can happen in different ways and one
such way may be the overlapping binding sites on the promoter (Fig.
1). Therefore, the activator and repressor molecules cannot bind the
promoter simultaneously, rather they compete for their binding sites
to regulate gene transcription. This mechanism of transcriptional
regulation is represented by a simple reaction scheme (Fig. 2) where
a gene can be in three possible states: $G_{1}$, $G_{2}$ and $G_{3}$.
$G_{2}$ is the unregulated state and $G_{1}$ ($G_{3}$) is the repressed
(activated) state of the gene. The unregulated state of the gene which
is achieved when both the sites are empty. Activator (repressor) molecules,
on binding its specific site, help in transition from the unregulated
state $G_{2}$ to the active (repressed) state $G_{3}$ ($G_{1}$)
of the gene. There are random transitions taking place between the
three states of the gene. Activator and repressor molecules compete
for the state $G_{2}$ to take control of the network. If activator
molecule wins, the gene turns into active state and protein synthesis
occurs with rate constant $J_{p}$. Protein production does not take
place from the unregulated ($G_{2}$) and repressed ($G_{1}$) states
of the gene. Degradation of proteins occur with rate constant $k_{p}$
and this event is independent of the states of the gene. Here transcription
and translation are combined together into a single step as done in
earlier studies \cite{Kepler,Karmakar}. The stochastic transition
from $G_{2}$ to $G_{3}$ occurs with rate constant $k_{a}$ and that
from $G_{2}$ to $G_{1}$ with $k_{2}$ (Fig. 2). The rate constants
$k_{a}$ and $k_{2}$ are the functions of activator and repressor
molecules respectively. Thus, in absence of repressor (activator)
molecules the transition from $G_{2}$ to $G_{1}$ ($G_{3}$) is not
possible at all. The assumption that there can be three possible states
of the gene provides the basis for the minimal model of the activator-repressor
system.

\begin{figure}
\begin{centering}
\includegraphics[width=2in]{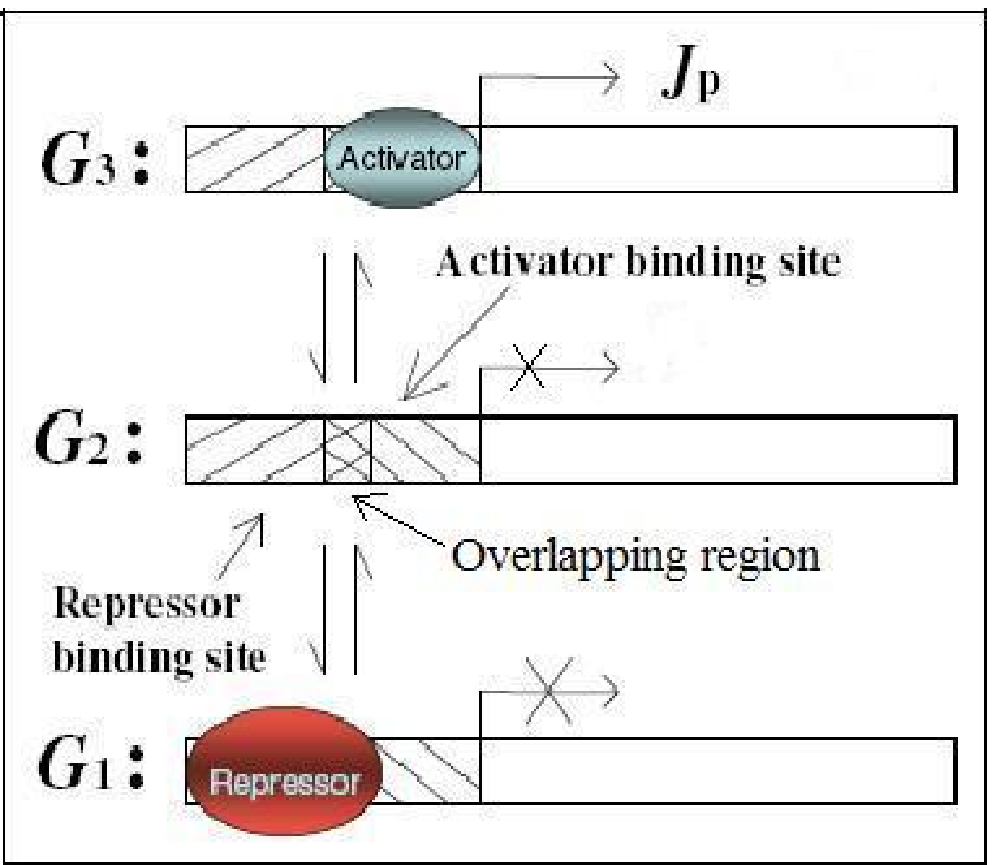}
\par\end{centering}

\vspace{0.25cm}

FIG. 1. Schematic diagram of transcriptional regulation by activator
and repressor molecules where both the molecules compete for their
respective binding site.
\end{figure}

\begin{figure}
\begin{centering}
\includegraphics[width=3in]{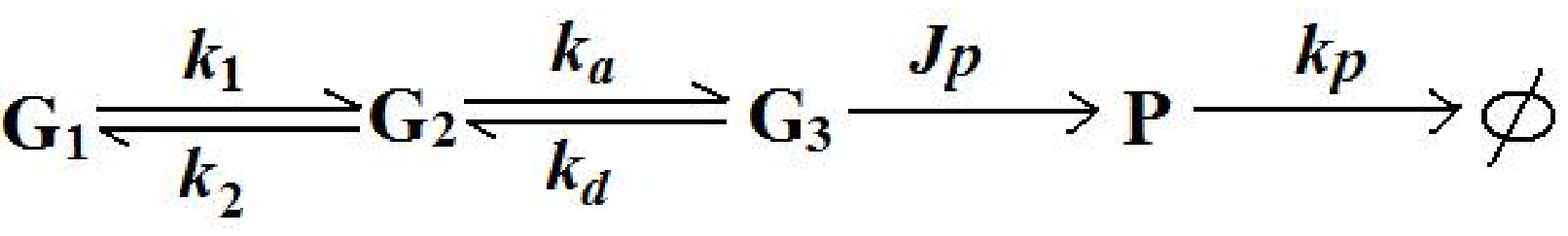}
\par\end{centering}

FIG. 2. Reaction scheme with the three states of the gene: repressed
($G_{1}$), unregulated ($G_{2}$) and activated ($G_{3}$). From
the activated state $G_{3}$ proteins are synthesized with rate constant
$J_{p}$. 
\end{figure}

Let $p_{i}(n,t)$ ($i=1,2,3$) be the probability that at time $t,$
the gene is in the $G_{i}$ state with $n$ number of protein molecules
in the system. The Master equations for the biochemical reactions
corresponding to the Fig. 2 are given by

\begin{equation}
\frac{\partial p_{1}(n,t)}{\partial t}=k_{2\,}p_{2}(n,t)-k_{1\,}p_{1}(n,t)+k_{p}[(n+1)p_{1}(n+1,t)-n\, p_{1}(n,t)]\label{eq:1}\end{equation}

\[
\frac{\partial p_{2}(n,t)}{\partial t}=k_{1\,}p_{1}(n,t)+k_{d\,}p_{3}(n,t)-k_{2\,}p_{2}(n,t)-k_{a\,}p_{2}(n,t)+J_{0}[p_{2}(n-1,t)-p_{2}(n,t)]\]

\begin{equation}
+k_{p}[(n+1)p_{2}(n+1,t)-n\, p_{2}(n,t)]\label{eq:2}\end{equation}
 \begin{equation}
\frac{\partial p_{3}(n,t)}{\partial t}=k_{a\,}p_{2}(n,t)-k_{d}\, p_{3}(n,t)+J_{p}[p_{3}(n-1,t)-p_{3}(n,t)]+k_{p}[(n+1)p_{3}(n+1,t)-n\, p_{3}(n,t)]\label{eq:3}\end{equation}

Now the standard approach of the theory of stochastic processes will
be used to determine the steady-state probability density function
for protein levels \cite{van kampen}. The generating functions are
defined as \begin{equation}
F_{1}(z,\, t)=\sum_{n}z^{n}\, p_{1}(n,t),\:\; F_{2}(z,\, t)=\sum_{n}z^{n}\, p_{2}(n,t),\;\;{\textstyle F_{3}(z,\, t)=\sum_{n}z^{n}\, p_{3}(n,t)\textrm{ and}}\; F(z,t)=\sum_{n}z^{n}\, p(n,t)\label{eq:4}\end{equation}

where\begin{equation}
\begin{array}{c}
F(z,t)=F_{1}(z,t)+F_{2}(z,t)+F_{3}(z,t)\\
p(n,t)=p_{1}(n,t)+p_{2}(n,t)+p_{3}(n,t)\end{array}\label{eq:5}\end{equation}

where $F(z,t)$ and $p(n,t)$ are the total generating function and
total probability density function respectively.

In terms of the generating functions (\ref{eq:4}), Eqs. (\ref{eq:1}),
(\ref{eq:2}) and (\ref{eq:3}) can be written as \begin{equation}
\frac{\partial F_{1}(z,t)}{\partial t}=k_{2}F_{2}(z,t)-k_{1}F_{1}(z,t)+k_{p}(1-z)\frac{\partial F_{1}(z,t)}{\partial z}\label{eq:6}\end{equation}

\begin{equation}
\frac{\partial F_{2}(z,t)}{\partial t}=k_{1}F_{1}(z,t)+k_{d}F_{3}(z,t)-k_{2}F_{2}(z,t)-k_{a}F_{2}(z,t)+k_{p}(1-z)\frac{\partial F_{2}(z,t)}{\partial z}\label{eq:7}\end{equation}
 \begin{equation}
\frac{\partial F_{3}(z,t)}{\partial t}=k_{a}F_{2}(z,t)-k_{d}F_{3}(z,t)+J_{p}(z-1)F_{3}(z,t)+k_{p}(1-z)\frac{\partial F_{3}(z,t)}{\partial z}\label{eq:8}\end{equation}

\begin{flushleft}
In the steady state (${\displaystyle \frac{\partial F_{i}}{\partial t}}=0$,
$i=1,2,3$), addition of Eqs. (\ref{eq:6}), (\ref{eq:7}) and (\ref{eq:8})
results 
\par\end{flushleft}

\begin{equation}
J_{p}F_{3}(z)=k_{p}\frac{\partial F(z)}{\partial z}\label{eq:9}\end{equation}

With the help of the Eqs. (\ref{eq:5}), (\ref{eq:6}), (\ref{eq:8})
and (\ref{eq:9}), $F_{1}(z)$ and $F_{2}(z)$ can be expressed in
terms of $F(z).$ Then, in terms of the generating function $F(z)$,
the Eqs. (\ref{eq:6}), (\ref{eq:7}) and (\ref{eq:8}) can be written
as

\begin{equation}
(z-1)^{2}F^{'''}(z)+\{a_{1}(z-1)-b_{1}(z-1)^{2}\}F^{''}(z)+\{a_{2}-b_{1}b_{2}(z-1)\}F^{'}(z)-b_{1}a_{3}F(z)=0\label{eq:10}\end{equation}

where $a_{1}=(1+s_{1}+s_{2}+s_{a}+s_{d})$, $b_{1}=J_{p}/k_{p}$,
$a_{2}=s_{1}s_{a}+s_{1}s_{d}+s_{2}s_{d}$, $b_{2}=a_{1}-s_{d}$, $a_{3}=s_{1}s_{a}$,
$s_{1}=k_{1}/k_{p},\: s_{2}=k_{2}/k_{p},\: s_{a}=k_{a}/k_{p}$ and
$\: s_{d}=k_{d}/k_{p}$.

\noindent The solution of the Eq. (\ref{eq:10}) is a generalized
hypergeometric function and is given by

\begin{equation}
F(z)=C\,_{p}F_{q}[g_{1}-g_{2};\, g_{1}+g_{2};\, h_{1}-h_{2};\, h_{1}+h_{2};\: b_{1}(z-1)]\label{eq:11}\end{equation}

where $g_{1}=-\frac{1}{2}+\frac{b_{2}}{2}$, $g_{2}=\frac{1}{2}\sqrt{(b_{2}-1)^{2}-4a_{3}},$
$h_{1}=-\frac{1}{2}+\frac{a_{1}}{2}$ , $h_{2}=\frac{1}{2}\sqrt{(a_{1}-1)^{2}-4a_{2}}$,
$C$ is the normalization constant and $_{p}F_{q}(a,b,c,d)$ is the
generalized hypergeometric function (GHF). The normalization constant
can be determined easily from the condition $F(1)=1.$

\begin{figure}
\begin{centering}
\includegraphics[width=3in]{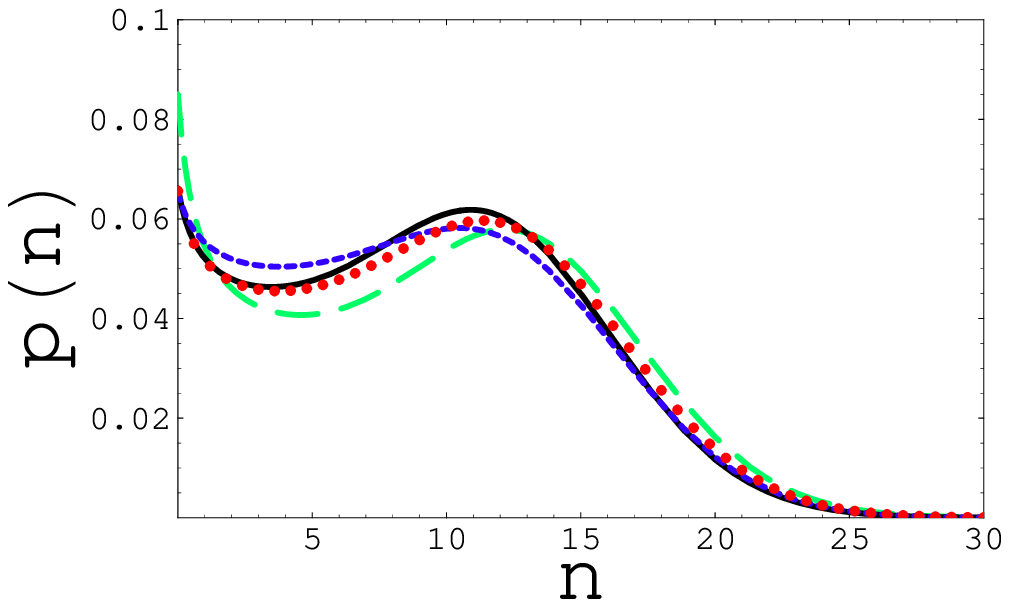}
\par\end{centering}

FIG. 3. Plot of $p(n)$ versus $n$ for the activator-repressor system
for \textbf{$b_{1}=16$} and four different sets of parameter values:
long dashed curve: $s_{1}=1,$ $s_{2}=6$, $s_{a}=10,$ $s_{d}=1$,
for solid curve: $s_{1}=1,$ $s_{2}=4$, $s_{a}=13,$ $s_{d}=2$,
for short dashed curve: $s_{1}=2,$ $s_{2}=6$, $s_{a}=5,$ $s_{d}=1$
and for dotted curve: $s_{1}=1.25,$ $s_{2}=6,$ $s_{a}=10,$ $s_{d}=1.25.$

\end{figure}

\begin{figure}

\begin{centering}
\includegraphics[width=3in]{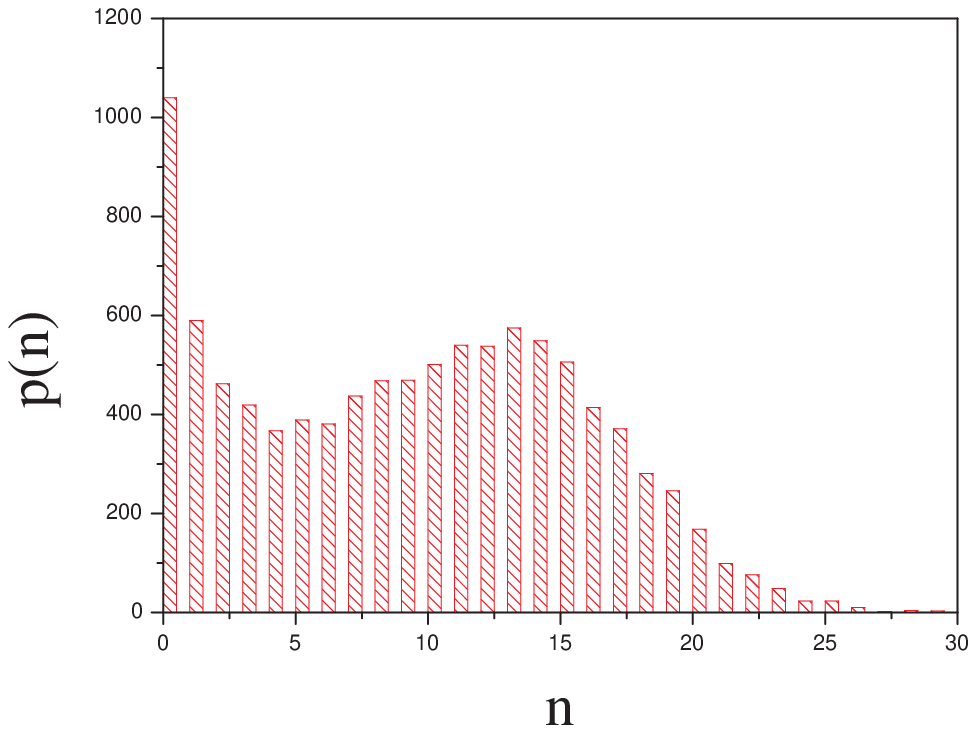}
\par\end{centering}

FIG. 4. Plot of $p(n)$ versus $n$ obtained from stochastic simulation
using Gillespie algorithm with the rate constants $k_{1}=1,$ $k_{2}=6$,
$k_{a}=10,$ $k_{d}=1$, $J_{p}=16$ and $k_{p}=1$ (same as the long
dashed curve of Fig. 3). For $k_{p}=1$, $s_{i}=k_{i}\,(i=1,2,a,d)$
and $b_{1}=J_{p}$.

\end{figure}

Differentiating Eq. (\ref{eq:11}) $n$ times w.r.t. $z$ at $z=0$,
one can easily obtain the expression for the steady-state probability
density function $p(n)$ as

\begin{equation}
p(n)=C\,\frac{b_{1}^{n}\,\Gamma(g_{1}+n)\,\Gamma(g_{2}+n)\,\Gamma(h_{1})\,\Gamma(h_{2})}{n!\,\Gamma(h_{1}+n)\,\Gamma(h_{2}+n)\,\Gamma(g_{1})\,\Gamma(g_{2})}\,_{p}F_{q}(g_{1}+n;\, g_{2}+n;\, h_{1}+n;\, h_{2}+n;\,-b_{1})\label{eq:12}\end{equation}
 The plot of $p(n)$ versus $n$ for different values of $s_{i\,}(i=1,2,a,d)$
with $b_{1}=16$ is shown in Fig. 3. Different curves in Fig. 3 show
that the distributions of protein levels are bimodal in different
parameter regions with $s_{i\,}(i=1,2,a,d)\geqslant1$. The binary
response can also be observed in a region of parameter values with
$s_{i\,}(i=1,2,a,d)<1$ (not shown). The binary responses in the activator-repressor
system cannot be observed for $s_{1},\, s_{d}\geqslant2$ (simultaneously).
Figure 4 shows the binary response in protein levels obtained from
stochastic simulation using Gillespie algorithm \cite{Gillespie}
for the biochemical reactions shown in Fig. 2 for the rate constants
$k_{1}=1,$ $k_{2}=6$, $k_{a}=10,$ $k_{d}=1$, $J_{p}=16$ and $k_{p}=1$
(For $k_{p}=1$, $s_{i}=k_{i}\,(i=1,2,a,d)$ and $b_{1}=J_{p}$).

\noindent To understand the origin of bimodal distribution in protein
levels in the present scenario we calculate the components of probability
density function $p_{i}(n)(i=1,2,3)$ in the steady state. Using Eqs.
(\ref{eq:9}) and (\ref{eq:11}) one can easily obtain $p_{3}(n)$
and is given by\begin{equation}
p_{3}(n)=C\,\frac{b_{1}^{n}\,\Gamma(g_{1}+n+1)\,\Gamma(g_{2}+n+1)\,\Gamma(h_{1})\,\Gamma(h_{2})}{n!\,\Gamma(h_{1}+n+1)\,\Gamma(h_{2}+n+1)\,\Gamma(g_{1})\,\Gamma(g_{2})}\,_{p}F_{q}(g_{1}+n+1;\, g_{2}+n+1;\, h_{1}+n+1;\, h_{2}+n+1;\,-b_{1})\label{eq:13}\end{equation}

\noindent In the steady state, differentiating Eq. (\ref{eq:8}) $n$
times w. r. t. $z$ at $z=0$ we have \begin{eqnarray*}
 & p_{2}(n)= & \frac{(s_{d}+b_{1})}{s_{a}}p_{3}(n)-\frac{b_{1}^{n+1}\,\Gamma(g_{1}+n+2)\,\Gamma(g_{2}+n+2)\,\Gamma(h_{1})\,\Gamma(h_{2})}{k_{p}\, s_{a}\, n!\,\Gamma(h_{1}+n+2)\,\Gamma(h_{2}+n+2)\,\Gamma(g_{1})\,\Gamma(g_{2})}\times\end{eqnarray*}
 \begin{eqnarray}
\qquad\qquad\qquad & \qquad\qquad\qquad & \,_{p}F_{q}(g_{1}+n+2;\, g_{2}+n+2;\, h_{1}+n+2;\, h_{2}+n+2;\,-b_{1})\label{eq:14}\end{eqnarray}

\noindent From Eq. (\ref{eq:5}) we have

\begin{equation}
p_{1}(n)=p(n)-p_{2}(n)-p_{3}(n)\label{eq:15}\end{equation}

\noindent where $p_{2}(n)$ and $p_{3}(n)$ are obtained from Eqs.
(\ref{eq:13}) and (\ref{eq:14}) respectively.

\begin{figure}
\begin{centering}
\includegraphics[width=3in]{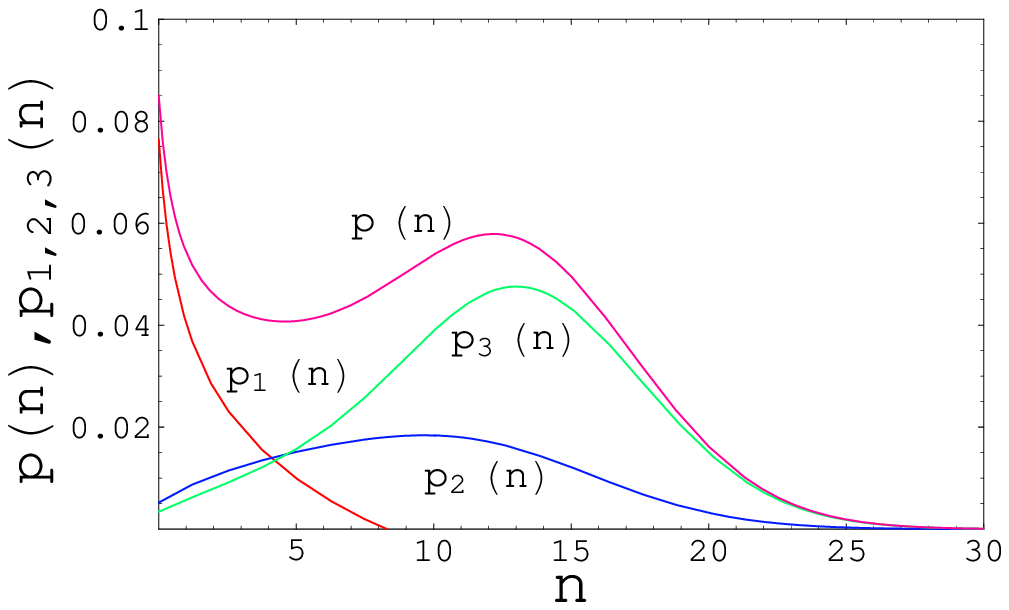}
\par\end{centering}

FIG. 5. Plot of $p(n)$ and $p_{i}(n)$ ($i=1,2,3$) versus $n$ for
the rate constants $s_{1}=1,$ $s_{2}=6$, $s_{a}=10,$ $s_{d}=1$
and $b_{1}=16$. The bimodal nature of the function $p(n)$ is the
resultant effect of three unimodal functions $p_{i}(n)(i=1,2,3)$.

\end{figure}

Figure 5 shows the plot of total and component probability density
functions $p(n)$ and $p_{i}(n)(i=1,2,3)$ respectively versus $n$,
the number of proteins, for the rate constants $s_{1}=1,$ $s_{2}=6$,
$s_{a}=10,$ $s_{d}=1$ and $b_{1}=16$ (same as the dotted curve
in Fig. 3). The bimodal distribution in protein levels is clearly
the resultant of three unimodal functions $p_{i}(n)$ $(i=1,2,3)$
(Fig. 5). This is also true for other bimodal curves in Fig. 3. From
the rate constants used to obtain the bimodal distributions in Fig.
3, it is clear that the probability of occurrence of the gene in the
$G_{1}$ and $G_{3}$ states are higher. Once it is in the $G_{1}$
state, the probability of transition to $G_{2}$ state is lower because
of the lower value of $s_{1}$. On the other hand, at any instant
of time, if the gene is in the $G_{2}$ state, there can be two possibilities:
gene can switch either to the active state ($G_{3}$) or to the repressed
state ($G_{1}$) depending on the amount of activators or repressors
present in the system since activator molecules modulate the transition
from $G_{2}$ to $G_{3}$ state and repressor molecules modulate the
transition from $G_{2}$ to $G_{1}$ state. Higher values of $s_{a}$
and $s_{2}$ make the $G_{3}$ and $G_{1}$ states more probable than
$G_{2}$ and due to the lower values of $s_{1}$ and $s_{d}$, the
gene spends most of the time either in the $G_{3}$ or $G_{1}$ state.
Once the gene is in the $G_{3}$ state, the transition to $G_{2}$
state is rare because of the lower value of $s_{d}$. From the $G_{3}$
state of the gene, proteins are synthesized with rate constant $J_{p}$
and there is enough time for the protein level to reach the steady
value. This gives rise to high protein level in single cell and the
peak in the distribution of protein level at higher value. This is
clearly observed in the curve for $p_{3}(n)$ in Fig. 5. Now if the
gene switches suddenly to $G_{2}$ state then protein level starts
to decrease. The protein level keeps on decreasing as long as the
gene is in the $G_{2}$ state or switches to the $G_{1}$ state. If
the gene switches to the $G_{1}$ state the protein level decreases
and reaches zero value. This gives rise to low/zero protein level
in a single cell and the peak in the distribution of protein level
occurs at low/zero value. This is observed in the curve for $p_{1}(n)$
in Fig. 5. From the $G_{2}$ state, the gene can also switch back
to the $G_{3}$ state and this causes a rise of protein level again
from an intermediate value. Therefore, there is a finite probability
to observe the protein level at the intermediate value. The curve
for $p_{2}(n)$ shows a finite value at the intermediate region of
protein level (Fig. 5).

Rossi et al. examined whether an interplay of transcription factors
can convert a graded to binary response in gene expression \cite{Rossi}.
They designed an experiment in which the ratio of activator and repressor
molecules that bind to the same promoter can be modulated by a single
inducer molecule dox. Furthermore, the activator and repressor molecules
bind the overlapping binding sites on the same promoter mutually exclusively.
They analyze the graded and binary responses to the inducer molecule
by flow cytometery in large population of individual cells. Three
different cell populations viz. a dox regulated repressor ({}``repressor
only''), a dox regulated activator ({}``activator only'') and both
({}``activator+repressor'') were generated in the experiment to
study the role of positive and negative transcription factors. The
flow cytometric analysis of the activator only and repressor only
cell populations revealed a graded response (unimodal distribution)
of GFP expression at all dox concentrations. The binary response (two
distinct sub-populations) was observed in cells containing both activator
and repressor molecules for a range of intermediate dox concentrations.
With increasing dox level, the increase in the number of cells with
maximal level of GFP and decrease in the number of cells with low
GFP level is observed. Therefore, an all-or-none (binary) response
to the inducer level is observed in the experiment of Rossi et al.
\cite{Rossi} when a combination of activator and repressor molecules
act on the same promoter mutually exclusively. Moreover, since either
factor independently produces a graded response, the binary response
observed in cells with both the regulatory molecules is not due to
a dominant effect of one factor over the other but rather to their
combined effect.

\begin{figure}
\begin{centering}
\includegraphics[width=1.8in]{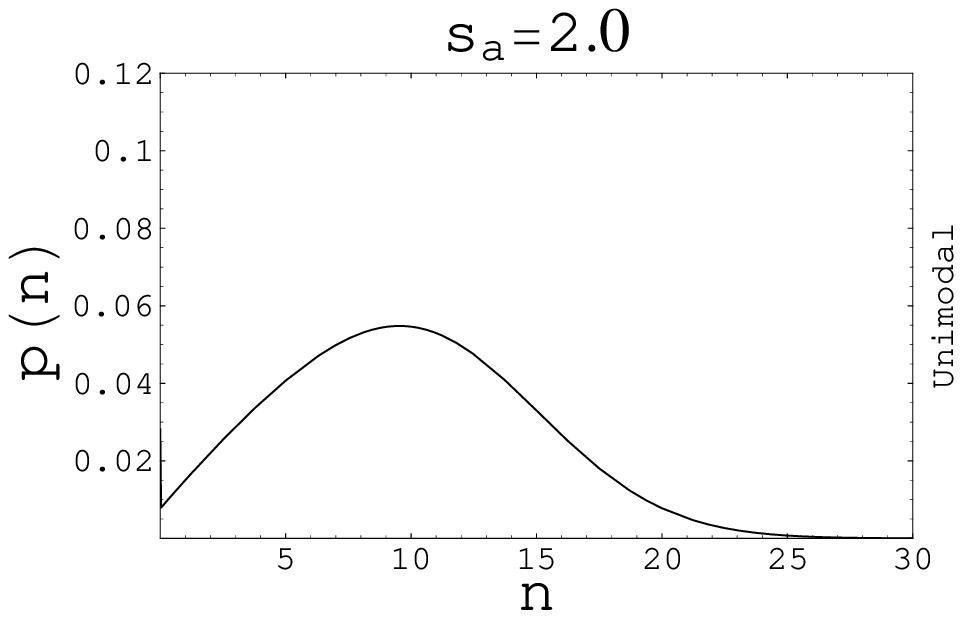} \includegraphics[width=1.8in]{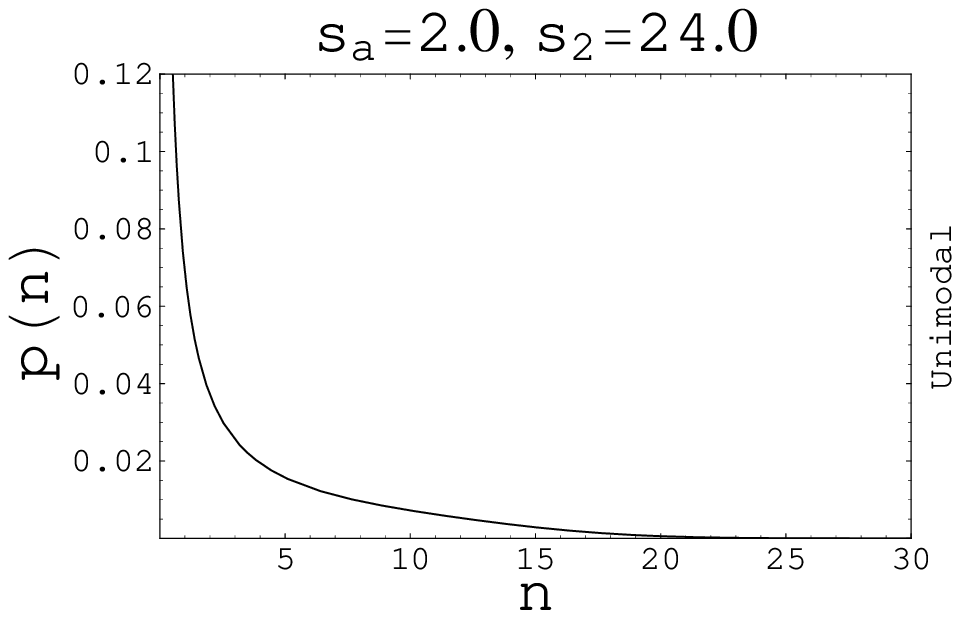}
\par\end{centering}

\begin{centering}
\includegraphics[width=1.8in]{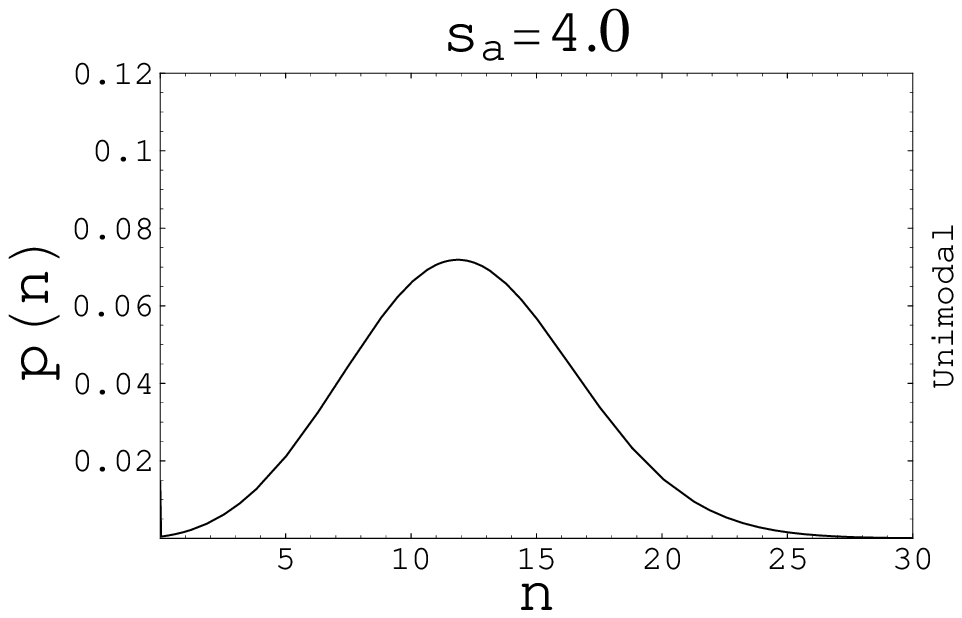} \includegraphics[width=1.8in]{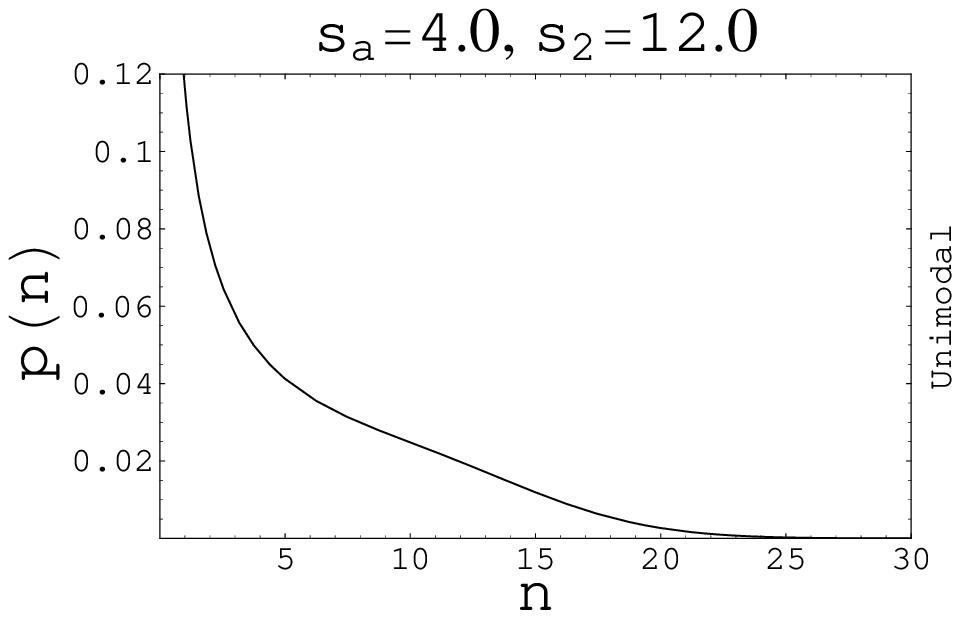}
\par\end{centering}

\begin{centering}
\includegraphics[width=1.8in]{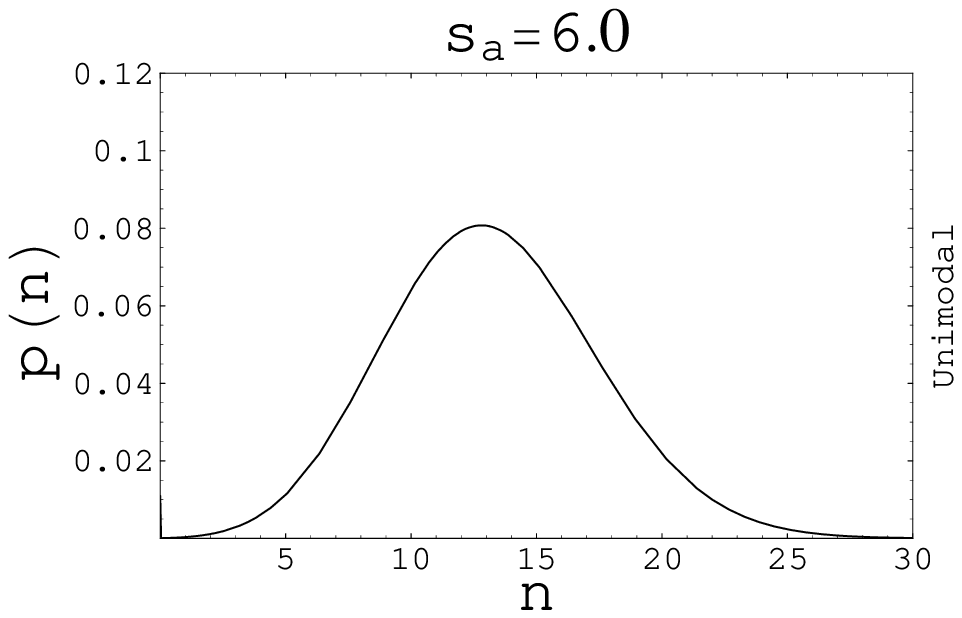} \includegraphics[width=1.8in]{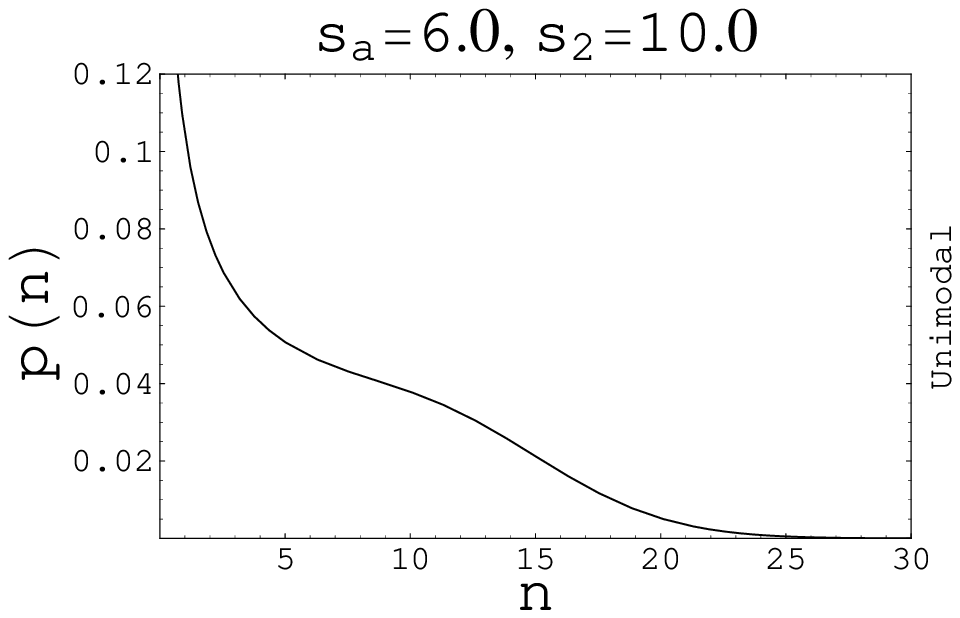}
\par\end{centering}

\begin{centering}
\includegraphics[width=1.8in]{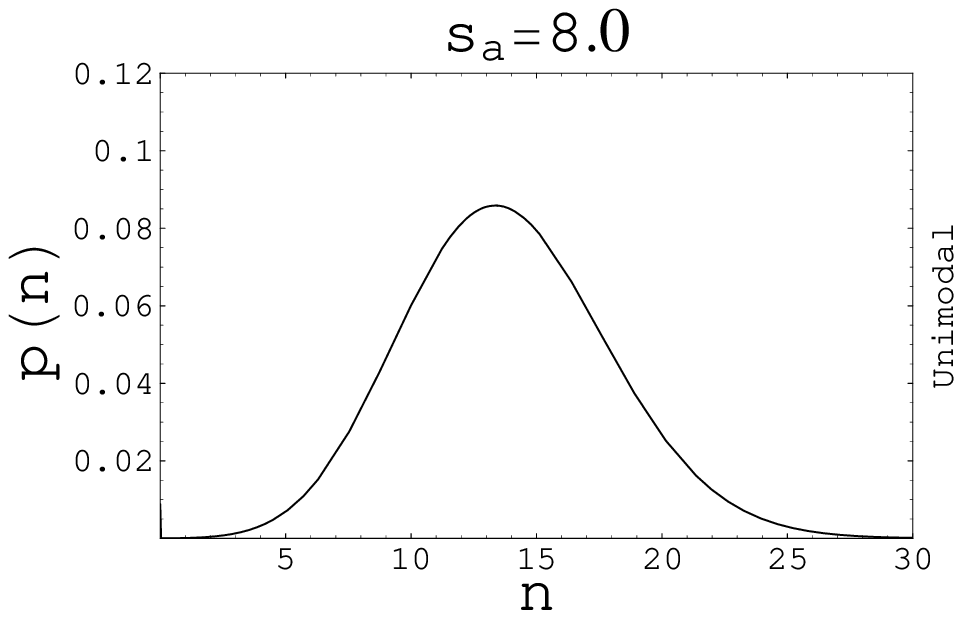} \includegraphics[width=1.8in]{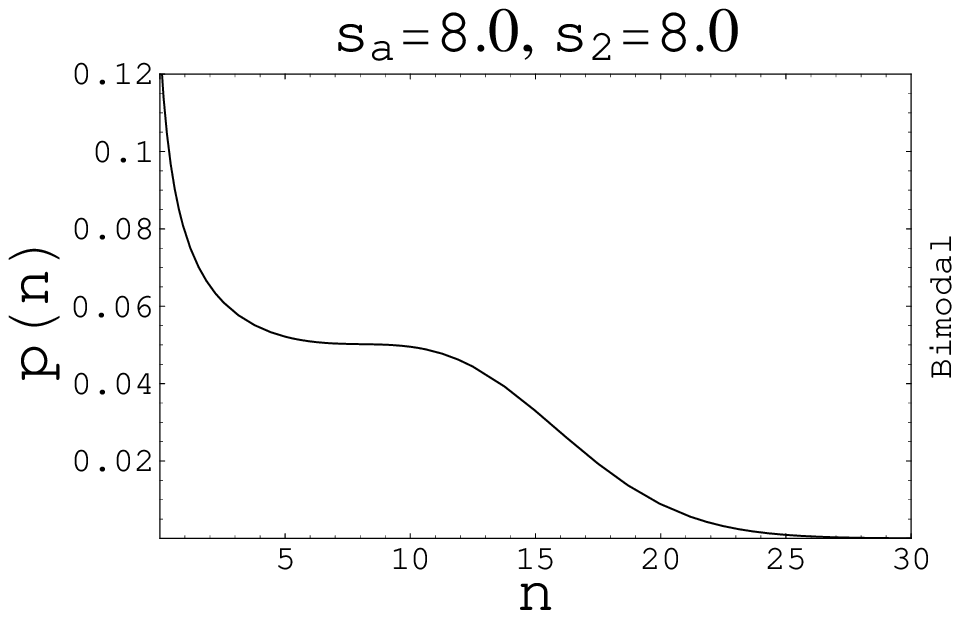}
\par\end{centering}

\begin{centering}
\includegraphics[width=1.8in]{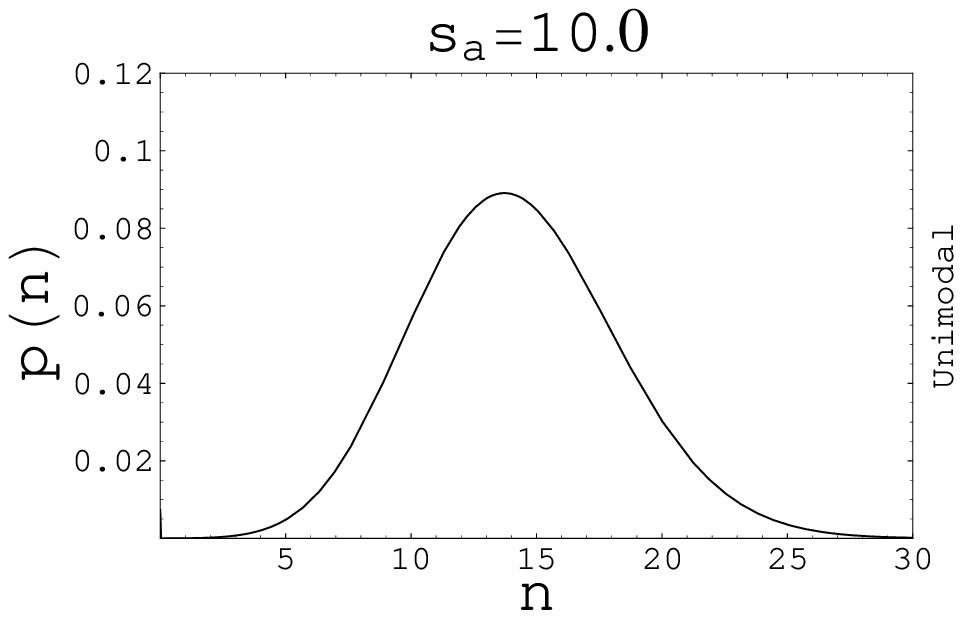} \includegraphics[width=1.8in]{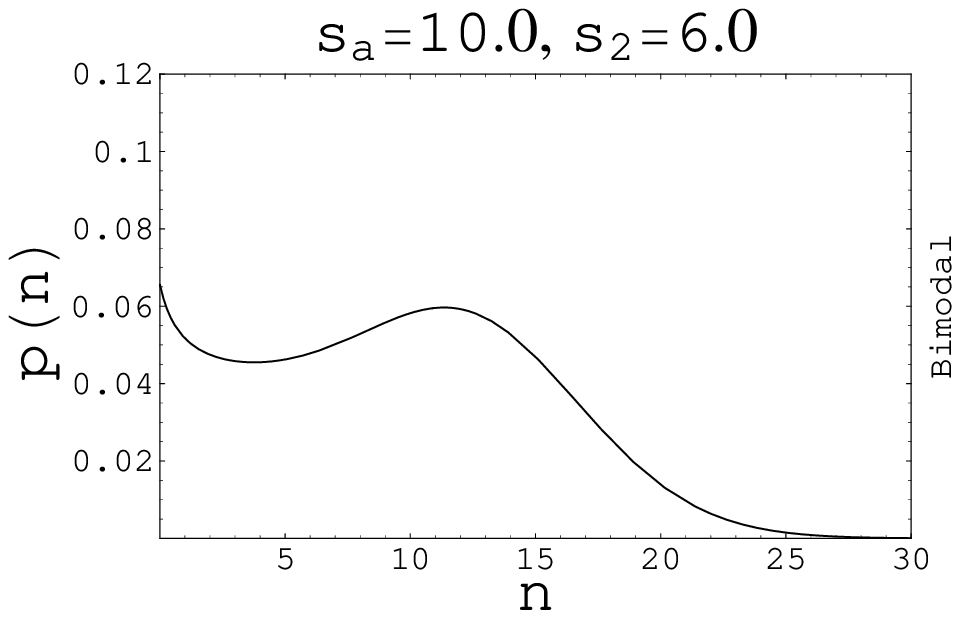}
\par\end{centering}

\begin{centering}
\includegraphics[width=1.8in]{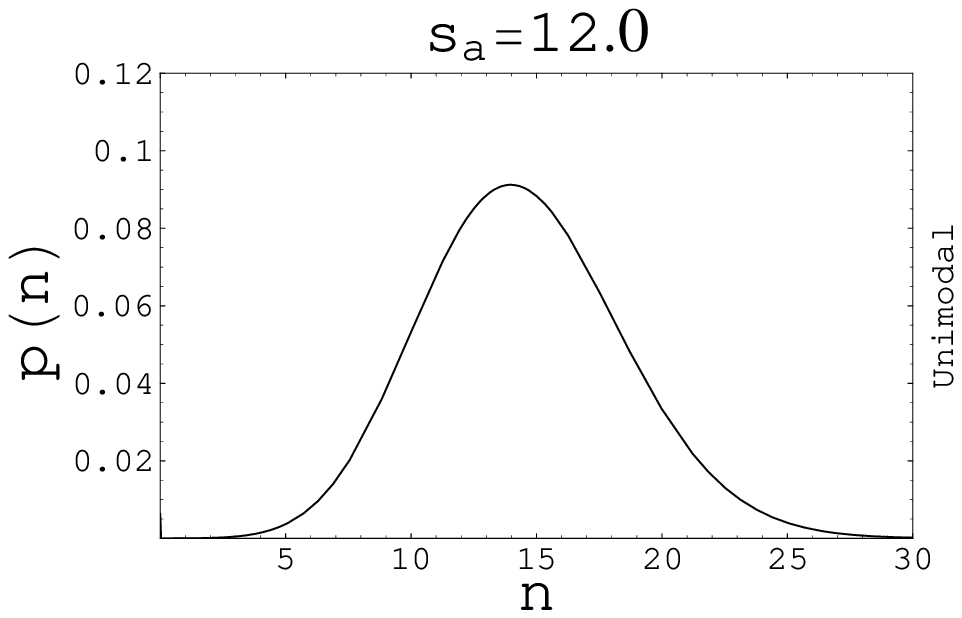} \includegraphics[width=1.8in]{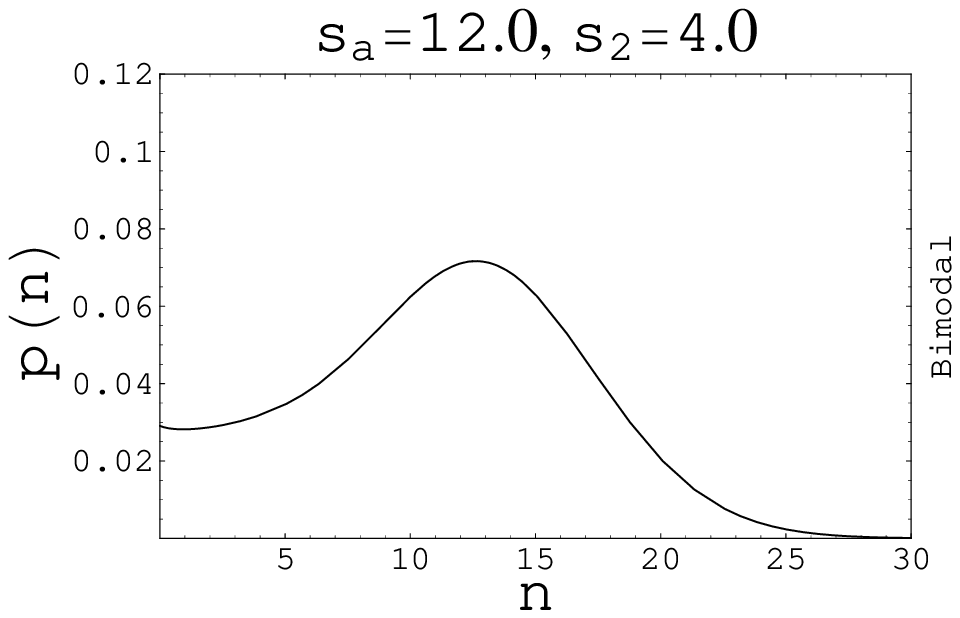}
\par\end{centering}

\begin{centering}
\includegraphics[width=1.8in]{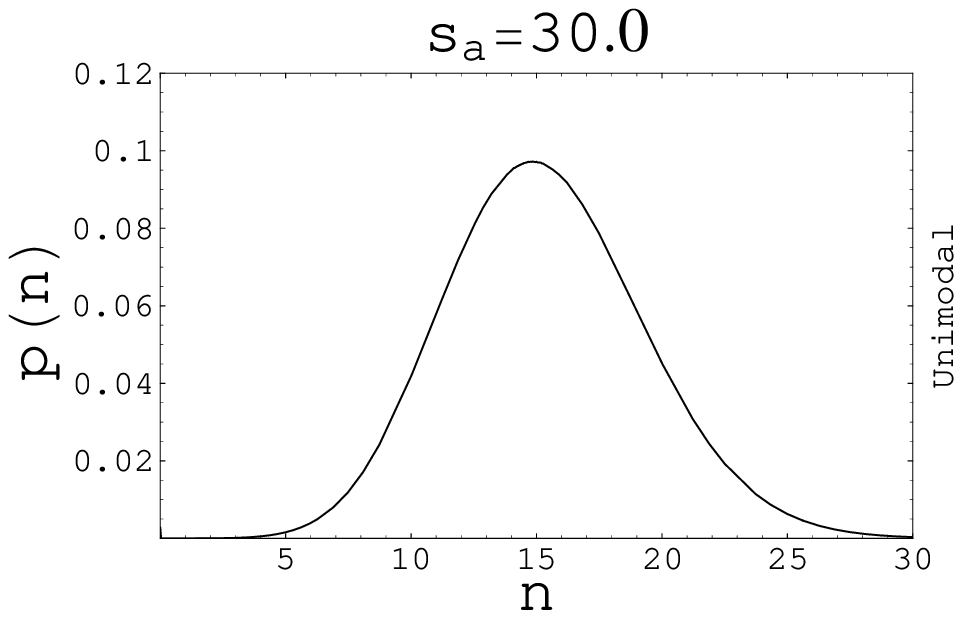} \includegraphics[width=1.8in]{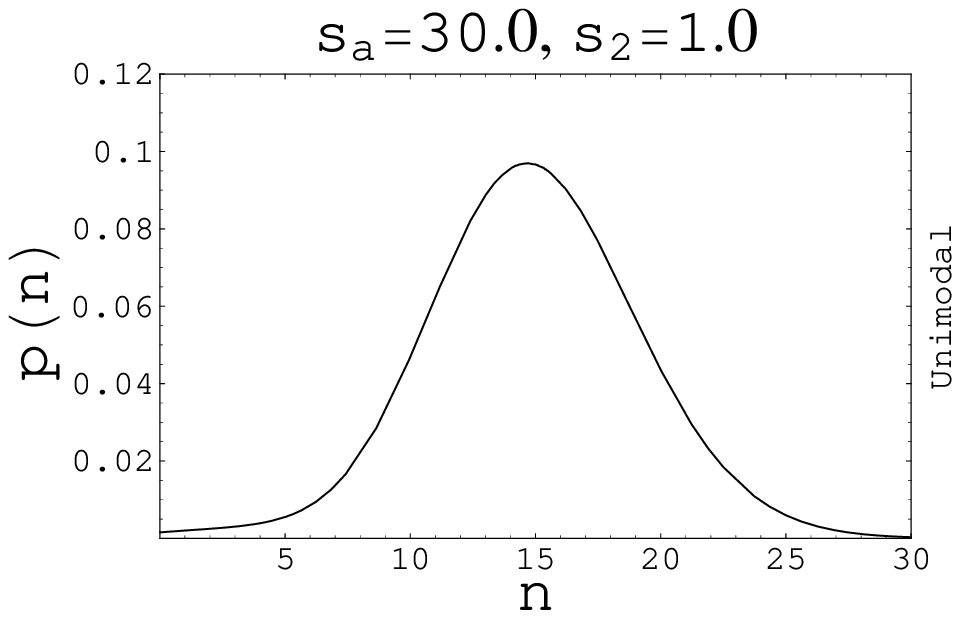}
\par\end{centering}

FIG. 6. Distribution of protein levels $p(n)$ versus $n$ in activator-only
system (left column) and activator-repressor system (right column)
for different level of inducer molecules i.e., for different values
of $s_{a}$ and $s_{2}$. For the activator-only system $s_{2}$ is
kept fixed at $1.25$ and $s_{a}$ is varied as mentioned on the top
of the figures (figures in the left column). For the activator-repressor
system $s_{1}$ and $s_{d}$ are kept fixed at $1.25$ and $1.25$
($s_{d}$ is same as in the activator-only system) respectively and
the different curves are drawn for different values of $s_{a}$ and
$s_{2}$ mentioned on the top of the figures (figures in the right
column). For all curves the relative transcription rate constant $b_{1}=16$.
\end{figure}

Our theoretical analysis of the activator-repressor system does not
explicitly include the activator and repressor numbers in the equations
but are included in the rate constants $s_{a}$ and $s_{2}$. The
rate constant $s_{a}$ increases with the increase in activator amount
and $s_{2}$ increases with the increase in repressor amount. Now,
let us assume that the numbers of both molecules can be controlled
by a single inducer molecule like dox, as in the experiment of Rossi
et al. \cite{Rossi}, so that $s_{a}$ increases and $s_{2}$ decreases
with the increase of dox. Depending on the presence of regulatory
molecules, the gene regulatory network can be divided into three categories:
activator-only system (i.e., only activator molecules regulate the
network), activator-repressor system (i.e., activators and repressors
both regulate the network) and repressor-only system (i.e., only repressor
molecules regulate the network). In presence of only activator molecules
the three-state gene activation process reduces to the two-state one.
Random switching then takes place only between $G_{2}$ and $G_{3}$
states. With the two-state gene activation process, the graded and
binary responses are observed for $s_{a},s_{d}>1$ and $s_{a},s_{d}<1$
respectively \cite{Karmakar}. In one hand, with $s_{d}>1$ if $s_{a}$
is varied from low to high value then unimodal responses are observed.
On the other hand, with $s_{d}<1$, if $s_{a}$ is varied from low
to high value then first unimodal (for $s_{a}<s_{d}$), then bimodal
(for $s_{a}\simeq s_{d}$) and then again unimodal (for $s_{a}>s_{d}$)
responses are observed \cite{Karmakar}. Rossi et al. observed graded
responses in activator-only system at all levels of inducer. To reproduce
the experimental observations of Rossi et al. we choose the parameter
region $s_{a},s_{d}>1$. Let us assume that initially there are only
activator molecules (with low copy number) activating the gene transcription
and $s_{d}$ is fixed at $1.25$. Now, with the gradual increase of
inducer molecules dox in the system, $s_{a}$ increases and the mean
protein level also increases gradually. The probability distributions
always remain graded (curves in the left column of Fig. 6) because
the values of $s_{a}$ and $s_{d}$ satisfy the condition of unimodal/graded
response ($s_{a},s_{d}>1$) for all values of dox \cite{Karmakar}\emph{.}
Let us now consider the same regulatory network (same $s_{d}$) but
with repressor molecules also present in the system. The gene can
now switch between all three possible states and both the molecules
compete for their binding site to take control of the gene transcription.
Let us assume that initially there are large number of repressors
i.e., $s_{2}$ is large and small number of activators i.e., $s_{a}$
is low. With the gradual increase of dox molecules in the system,
$s_{2}$ decreases and $s_{a}$ increases gradually and simultaneously
i.e., inhibition effect decreases and activation effect increases
simultaneously. This causes the conversion of unimodal (for low dox
i.e., low $s_{a}$ and high $s_{2}$) to bimodal (for intermediate
dox i.e., intermediate $s_{a}$ and $s_{2}$) and then again unimodal
(high dox i.e., high $s_{a}$ and low $s_{2}$) distribution of protein
levels (right column of Fig. 6). The gradual increase in the inducer
level causes a discontinuous change in the mean protein level. Therefore,
the response is bimodal/binary as the mean protein level is not a
continuous function of inducer but has only low and high values. These
results (Fig. 6) are in qualitative agreement with the experimental
observations of Rossi et al. \cite{Rossi} for activator-only and
activator-repressor systems.

Rossi et al. \cite{Rossi} also observed the graded response when
only repressor molecules regulate the gene transcription. To reproduce
the experimental observation for repressor-only case in the present
scenario, one has to consider the basal rate of protein synthesis
from the unregulated state of the gene ($G_{2}$). With the basal
rate of protein synthesis, say $J_{0}$ ($J_{0}<J_{p}$), from the
$G_{2}$ state, the generation of graded response for repressor-only
case is quite similar to that of the activator-only case discussed
above. In the presence of only repressor molecules in the system,
the three-state gene activation process reduces to the two-state one
i.e., the gene can switch randomly only between $G_{1}$ and $G_{2}$
states. Repressor molecules help in transition from $G_{2}$ to $G_{1}$
state. With the finite basal rate of protein synthesis from the $G_{2}$
state the response will be graded for $s_{1}>1$ and for all values
of $s_{2}$ \cite{Karmakar}\emph{.} The initial value of the rate
constant $s_{2}$ is large due to the presence of large number of
repressor molecules in the repressor-only system, the response in
this case will be unimodal since $s_{1}=1.25$ and $s_{2}$ is large.
Now with the gradual increase of dox concentration, the rate constant
$s_{2}$ decreases from a high to a low value but the response still
remains graded due to $s_{1}$ being greater than one ($s_{1}=1.25$).
Therefore, with the basal rate of protein synthesis from the unregulated
state the graded response can also be observed for repressor-only
system. But with the basal rate of protein synthesis from the state
$G_{2}$ the derivation of exact analytical expression for the probability
density function of protein levels for the activator-repressor system
is very difficult. Though, with the help of stochastic simulation
using Gillespie algorithm, it can be shown that the finite basal rate
of protein synthesis from the state $G_{2}$ does not change our results
qualitatively. The qualitative nature of the curves drawn in Figs.
3 and 6 will remain unchanged with the basal rate of gene expression
from $G_{2}$ taken into account.

\vspace{0.5cm}

\textbf{\large III. DISCUSSION}{\large \par}

\vspace{0.5cm}

In this paper, we have studied an gene regulatory network where the
positive and negative transcription factors regulate the gene transcription
mutually exclusively. Both the molecules compete for their respective
binding sites on the DNA to take control of the network (Fig. 1).
The activator-repressor system is represented by a simple stochastic
model where gene can be in three possible states viz. inactive/repressed,
unregulated and active. An exact analytical expression for the probability
density function of the protein levels in the steady state is derived
and is a generalized hypergeometric function (GHF) (Eq. (\ref{eq:12})).
From the GHF, the bimodal distribution in protein levels is observed
in a wide region of the parameter values. From the theoretical analysis,
the experimental observation of Rossi et al. (i.e., the regulation
only by activator molecules produce the graded response in the protein
levels whereas binary responses are observed when both the activator
and repressor molecules regulate the gene transcription by binding
the promoter mutually exclusively) can be reproduced very easily.
Here we have considered only the parameter region $s_{i}(i=1,2,a,d)\geq1$
(Fig. 3). The binary response in protein level is more prominent for
$s_{i}(i=1,2,a,d)<1$ (not shown). This region is excluded from the
present analysis because with $s_{i}(i=1,2,a,d)<1$, the binary response
can also be observed for two-state activator-only system \cite{Karmakar}.
But the experiment of Rossi et al. \cite{Rossi} observed only graded
response when only activator molecules regulate the gene transcription.
This experimental observation along with the theoretical prediction
of graded response in gene expression \cite{Karmakar} helps us to
choose the parameter region for theoretical analysis. Rossi et al.
observed the graded response also for repressor-only system. Here
we have not considered the repressor-only case because this requires
a basal rate of protein synthesis from the unregulated ($G_{2}$)
state. The basal rate of protein synthesis from the unregulated state
brings difficulties in the analytical tractability of the model. In
the presence of the basal rate of protein synthesis from the unregulated
state, it is very difficult to express the components of the generating
functions $F_{i}(z)$ ($i=1,2,3$) in terms of the total generating
function $F(z)$ and therefore the Chemical Master Equations (CME)
cannot be expressed by a single differential equation like Eq. (\ref{eq:10}).
Again, the reduction of the CME into a single differential equation
does not lead to the exact solution of CME because of the unavailability
of the analytical solution of the higher order differential equation.
This shows the limited scope and applicability of the generating function
technique used here to solve the CME. The difficulty increases when
the regulatory networks consist of nonlinear feedback loops. 

In the present analysis of activator-repressor system, we have combined
the transcription and translation into a single step process. In the
process of transcription mRNAs are produced from the active gene and
then mRNAs are translated into proteins. Therefore, the steady state
probability distribution (Eq. \ref{eq:12}) derived here gives the
correct description for mRNAs. The distribution in protein levels
does not follow the bimodal mRNA distribution when the protein lifetime
is longer than that of mRNA \cite{Jayaprakash}. Despite the above
limitations of the stochastic model, it contains important features
necessary for an explanation of the binary response in an activator-repressor
system and is graded in activator only system as observed in experiment
\cite{Rossi}. The exact analytical result with three gene states
is important and useful specially in the eukaryotic system. The gene
activation of the complex eukaryotic system consists of many unknown
number of rate limiting steps (chromatin remodeling, assembly of preinitiation
complex etc.). The simplification of the complex gene activation process
by the two-state one is the first approximation of the complicated
biological process. The 'three-state' assumption may be considered
as the second approximation of the stochastic gene activation-deactivation
process.

The present analysis of the origin of binary responses in three-state
model may be helpful to explain the bimodal distribution in transcriptional
silencing \cite{Xu}. In transcriptional silencing, Sir proteins (Sir
2-4) are the key structural components of silenced chromatin and under
their regulation the silencer can be in two possible states: repressed
and derepressed. The silencer helps to assemble the Sir protein complex.
This process of assembling is not a single step process but rather
consists of several reversible biochemical steps. The intermediate
steps between the repressed and derepressed states of the chromatin
make the effective rates of transitions very slow and these slow rate
of transitions ultimately may lead to the bimodal distribution of
protein levels from reporter gene.

\begin{center}
\textbf{Acknowledgement} 
\par\end{center}

This work is supported by the Minor Research Project Grant, UGC, India,
under Sanction No. F. PSW-001/07-08 (ERO).

\end{document}